\documentclass{article}
\usepackage{spconf,graphicx,tabularray,multirow}
\usepackage[fleqn]{amsmath}
\usepackage{hyperref}

\usepackage{enumitem}
\setlist{nosep, leftmargin=14pt}

\usepackage{mwe} 

\title{TC-DiffRecon: Texture coordination MRI reconstruction method based on diffusion model and modified MF-UNet method}
\name{
\parbox{\linewidth}{\centering
Chenyan Zhang$^1$, Yifei Chen$^1$, Zhenxiong Fan$^1$, Yiyu Huang$^1$, Wenchao Weng$^1$, \\
Ruiquan Ge$^{1,\star}$, Dong Zeng$^2$, Changmiao Wang$^{3,\star}$}
}

\address{$^1$Hangzhou Dianzi University, Hangzhou, China \\
$^2$Southern Medical University, Guangdong, China \\
$^3$Shenzhen Research Institute of Big Data, Shenzhen, China \\
\tt$\!\!\!\star$ Corresponding Author: gespring@hdu.edu.cn,cmwangalbert@gmail.com\\}
\begin{document}
%
\maketitle
\begin{abstract}
Recently, diffusion models have gained significant attention as a novel set of deep learning-based generative methods. These models attempt to sample data from a Gaussian distribution that adheres to a target distribution, and have been successfully adapted to the reconstruction of MRI data. However, as an unconditional generative model, the diffusion model typically disrupts image coordination because of the consistent projection of data introduced by conditional bootstrap. This often results in image fragmentation and incoherence. Furthermore, the inherent limitations of the diffusion model often lead to excessive smoothing of the generated images. In the same vein, some deep learning-based models often suffer from poor generalization performance, meaning their effectiveness is greatly affected by different acceleration factors. To address these challenges, we propose a novel diffusion model-based MRI reconstruction method, named TC-DiffRecon, which does not rely on a specific acceleration factor for training. We also suggest the incorporation of the MF-UNet module, designed to enhance the quality of MRI images generated by the model while mitigating the over-smoothing issue to a certain extent. During the image generation sampling process, we employ a novel TCKG module and a Coarse-to-Fine sampling scheme. These additions aim to harmonize image texture, expedite the sampling process,  while achieving data consistency. Our source code is available at \href{https://github.com/JustlfC03/TC-DiffRecon}{https://github.com/JustlfC03/TC-DiffRecon}.
\end{abstract}
\section{Introduction}
\label{sec:intro}

Magnetic Resonance Imaging (MRI) is a cutting-edge medical imaging technology extensively employed in clinical diagnosis and treatment. However, its lengthy acquisition process often results in image artifacts due to patient movement. To address this issue, it is common to undersample K-space to accelerate speed of MRI. MRI reconstruction seeks to mitigate this problem by utilizing the undersampled K-space to generate de-aliased MRI images \cite{c21madore2002using,c22yu2017deep}. This approach effectively preserves the original signal for clinical application, while also reducing the acquisition time of MRI images.

The remarkable success of deep learning across various research domains has instigated the utilization of multiple deep learning-based models in MRI reconstruction. These models include convolutional neural networks \cite{c4wang2023multimodal}, graph neural networks \cite{c7markova2022global}, recurrent neural networks \cite{c5he2016deep}, and transformers \cite{c8montana2022towards}. They have demonstrated exceptional performance in reconstructing high-quality images at high acceleration factors (AF).

In recent years, diffusion modeling\cite{c10ho2020denoising, c11nichol2021improved} has emerged as one of the most promising generative models. It consists of two basic processes forward process to perform noise addition and sampling process to gradually remove the noise from the image. Diffusion models have found extensive application in MRI reconstruction. For instance, Peng et al.~\cite{c16peng2022towards} proposed a DDPM-based MRI reconstruction method, termed DiffuseRecon, which exhibited superior reconstruction results for different AF when trained on fully sampled MRI images. Chung et al.~\cite{c14chung2022score} proposed a fractional-based MRI reconstruction model, which utilized a numerical SDE solver and completed the reconstruction task iteratively through data consistency steps. Güngör et al.~\cite{c15gungor2023adaptive} introduced AdaDiff to enhance the model's reconstruction performance during the inference stage. Nevertheless, none of them consider the inherent flaws present in the original U-Net of the diffusion model, which leads to excessive smoothing of the sampled images as well as poor imaging quality. In addition, since the diffusion model is an unconditional generative model, the data-consistent projection injected by the existing diffusion model-based MRI reconstruction methods to achieve conditional guidance usually impairs the image coherence.

In this work, we propose a novel Multi-Free U-Net (MF-UNet)  for the TC-DiffRecon model for MRI reconstruction. The MF-UNet incorporates two uniquely designed modulation factors, which dynamically balance the contributions from the trunk and jump-connect features intrinsic to the U-Net architecture \cite{c20ronneberger2015u}. The first factor, known as the backbone feature factor, is capable of amplifying the primary backbone features, thereby enhancing the denoising process. However, our findings indicate that while the inclusion of the backbone feature scaling factor considerably improves the overall performance, it might lead to an over-smoothing effect on the resultant image texture. To mitigate this issue, we introduce the second factor - the skip feature scaling factor - which effectively alleviates the problem of texture over-smoothing.

Meanwhile, we introduce a novel Texture Coordination K-space Guidance (TCKG) module, which performs data consistency operations in K-space. By employing a texture coordination strategy, we eliminate texture inconsistency and screen fragmentation that could potentially arise from data consistency operations. Furthermore, we adopt a Coarse-to-Fine (C2F) sampling strategy to expedite the sampling process and further alleviate texture incoherence. Of particular note, our TC-DiffRecon model exhibits significant generalizability to MRI reconstruction of undersampled images across varying AF. We conduct experiments on the FastMRI dataset \cite{c17zbontar2018open} and the results indicate that our model outperforms the SoTA methods in terms of generalizability and image quality.

\section{Method}

\subsection{Model overview}

\begin{figure*}[htb]
\centerline{\includegraphics[width=0.8\textwidth]{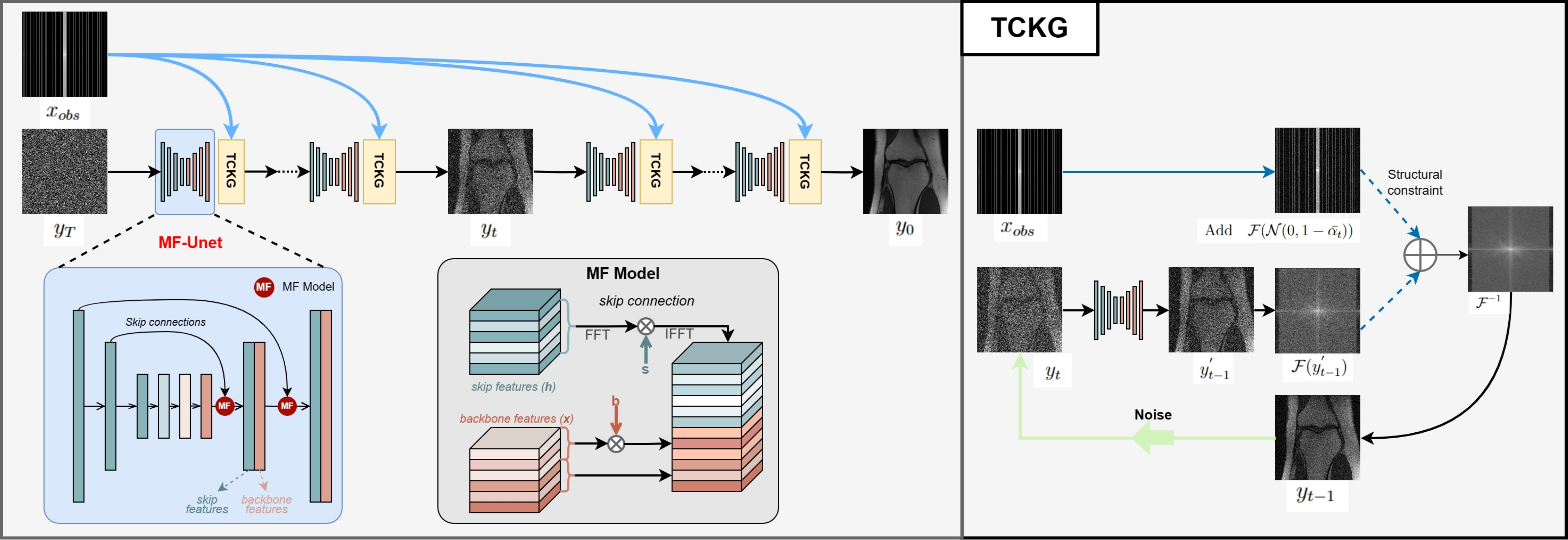}}
\caption{Overall structure of the TC-DiffRecon. It substitutes U-Net in the diffusion model with MF-UNet, which can dynamically adjust the weights of jump connections and backbone connections by two feature coordination factors. Within the TCKG module, the model progressively integrates K-space information into the denoising process in a texture-coordinated manner.}
\label{fig1}
\end{figure*}

The architecture of our proposed TC-DiffRecon model is illustrated in Fig. \ref{fig1}. The MF-UNet module is utilized to predict noise at the t-1 step. Concurrently, we employ the TCKG module for iterative adjustments to reconstruct high-fidelity target MRI images, conditioned on $x_{obs}$. In this process, we find that data consistency is ensured while maintaining the image texture harmonization.

\subsection{MF-UNet module}

For the $l$-th block of the decoder in U-Net, $x_l$ signifies the backbone features of the previous block's backbone, while $h_l$ represents the feature map propagated through corresponding skip connections. According to previous research \cite{c18si2023freeu}, it is suggested that the backbone features of the U-Net in the diffusion model contribute to denoising. Simultaneously, the skip connections introduce high-frequency features into the decoder, thereby accelerating the model's convergence to noise. However, this results in overlooking the fundamental backbone semantics, subsequently weakening the denoising capabilities of the backbone network. To harmonize these feature mappings, our MF-UNet employs two scalar factors, in addition to the multi-head attention mechanism - a backbone feature scaling factor bl for $x_l$, and a skip feature scaling factor $s_l$ for $h_l$. Specifically, the factor bl is designed to amplify the backbone feature map $x_l$, while the factor $s_l$ is purposed to attenuate the skip feature map $h_l$. For backbone features, we adaptively adjust the scaling factor based on the feature maps' average, $\overline{x}_l$:

\vspace{-1.5em}
\begin{equation}
\alpha _l=(b_l-1)\cdot {\overline{x}_l - Min(\overline{x}_l)\over{Max(\overline{x}_l)-Min(\overline{x}_l)}}+1,
\label{eq4}
\end{equation}
\vspace{-0.5em}
\begin{equation}
x_{l, i}^\prime = 
\begin{cases} 
  x_{l, i} \odot \alpha_l, & \text{if } i < C/2, \\  
  x_{l, i}, & \text{otherwise}, 
\end{cases}
\label{eq5}
\end{equation}
\vspace{-1em}

where $x_{l,i}$ represents the $i$-th channel of the feature mapping $x_l$, and C indicates the total number of channels in $x_l$. The backbone factor is represented by $\alpha_l$, while $b_l$ denotes a scalar constant. It should be noted, however, that scaling the backbone features may partially compromise the high-frequency details of the image during the denoising process. Therefore, to prevent the generation of excessively smooth textures in the synthesized image, we limit the scaling operation to half the channels of $x_l$, as outlined in Eq.(\ref{eq5}).

For the skip feature,  we additionally utilize spectral modulation in the Fourier domain. This approach aims to selectively diminish the low-frequency component of the skip feature. Our intention is to further alleviate the issue of overly smooth textures, a problem commonly caused by enhancement denoising, as will be illustrated in follows:

\vspace{-1.5em}
\begin{equation}
\begin{aligned}
    h_{l, i}^\prime &= \mathcal{F}^{-1}(\mathcal{F}(h_{l, i})\odot \beta_{l, i}), \\
    \beta_{l, i}(r) &= 
    \begin{cases} 
        s_l, & \text{if } r < r_{\text{thresh}}, \\  
        1, & \text{otherwise},
    \end{cases}
\end{aligned}
\label{eq6}
\end{equation}
\vspace{-1em}

where $\odot$ signifies pixel-level multiplication, while $\mathcal{F}$ and $\mathcal{F}^{-1}$ represent the Fourier transform and the inverse Fourier transform, respectively. The Fourier mask function, symbolized as $\beta_{l,i}$, specifies the range of the scale factor $s_l$ within the Fourier transformed image. Here, $r$ stands for the radius and $r_{thresh}$ is the threshold frequency.

\subsection{TCKG module}

\subsubsection{Data consistency strategy}

As the model acquires the prior knowledge of MRI images through training, and considering the distinctive physical properties of these images, we utilize K-space information for bootstrapping. This ensures the generated images uphold data consistency. We first obtain an initial unconditional output 
 $y_{t}^\prime$ according to the inverse process of the diffusion model. Subsequently, an under-sampled K-space image {$x_{obs}$ with a zero mean noise is added to simulate its diffusion state at step t and derive $x_{obs,t}$. Following this, an under-sampling mask $\mathcal{M}$ is employed to amalgamate the noisy observations $y_{t}^\prime$ mixed with the noise-added $x_{obs,t}$. This process is represented as follows:
\vspace{-1.5em}

\begin{gather}
y_t = \mathcal{F}^{-1}((1-\mathcal{M})\mathcal{F}y_t^\prime+\mathcal{M}  x_{\mathrm{obs},t} ),\label{eq8}\\
x_{\mathrm{obs},t}=x_{\mathrm{obs}}+ \mathcal{F}(\mathcal{N}(0,(1-\overline{\alpha_t})\mathrm{I})).
\end{gather}

The resultant $y_{t}$ is iteratively processed until it reaches $y_0$. Given that $x_{\text{obs}, 0} = x_{\text{obs}}$, $y_0$ retains a congruent K-space signal, achieving data consistency.

\subsubsection{Texture coordination strategy}

However, the data consistency operation can potentially result in textural discordance in the image, thereby compromising the quality of the generated image. At a specific time step t, the contents of $y_{t}^\prime$ might be incongruent with $\mathcal{F}^{-1}$$x_{obs,t}$, leading to discordance in the $y_{t-1}$ generated by Eq.(\ref{eq8}). During the denoising process at the subsequent time step t-1, the model attempts to rectify the discordance of $y_{t-1}$ to align with the $p_{\theta}(y_t)$ distribution. This process introduces new inconsistencies, preventing the model from converging, consequently resulting in a discordant and low-quality image. To address this important issue, as depicted in Fig.\ref{fig1}, we propose a texture coordination strategy: noise is reapplied to the $y_{t-1}$ generated at step t, creating a new $y_t$. Subsequently, denoising at step t is repeated K times, where K is a hyperparameter. 

\subsection{C2F sampling method}

Given the extended sampling time of the diffusion model, and with the aim of further addressing the texture incongruity in images and enhancing the quality of the generated images, we employ the C2F sampling method, which is inspired by related work \cite{c16peng2022towards}. Specifically, we accomplish this by compressing the T-step sampling process to uniformly spaced shorter schedules {T, T-k, ..., 1} \cite{c14chung2022score}, where $k>1$. Subsequently, we average multiple parallel samples to minimize the larger noise apparent in the image due to altering the diffusion step. Such a process also assists in mitigating texture incoherence introduced due to data consistency operations, thereby improving semantic accuracy. As represented in Fig.\ref{fig2}, C2F sampling generates N instances of $y_t$ $\sim$ N (0, I) and denoises each at the $T/k$ steps according to the new sampling step. The results of the noise are averaged to obtain $y_0^{avg}$. Ultimately, $y_0^{avg}$ is refined through an additional $T_{refine}$ step, which assists in eliminating ambiguity introduced by the sample averaging process, leading to more realistic reconstruction results. For data consistency, during the refinement step, the TCKG module directly utilizes $x_{obs}$ as input for data consistency.

\begin{figure*}
\centerline{\includegraphics[width=0.8\textwidth]{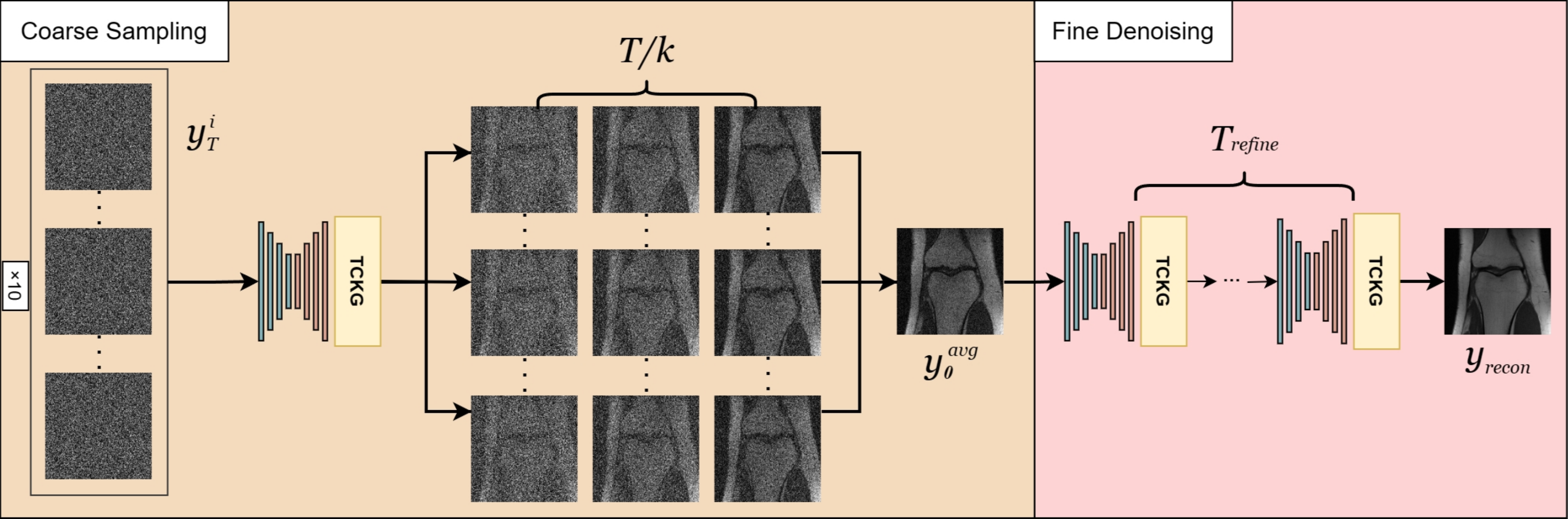}}
\caption{C2F Sampling Process. C2F sampling methods produce multiple samples following the T/k steps of denoising. These samples are subsequently averaged over $y_0^{avg}$, followed by iterative refinement through $T_{refine}$ steps.}
\label{fig2}
\end{figure*}

\section{Experiment Result}
\subsection{Dataset and  implementation details}

Our experiments are conducted using the FastMRI dataset \cite{c17zbontar2018open}, which includes K-space data for MRI. We utilized single-coil knee MRI scans from FastMRI, comprising 1172 subjects with approximately 35 slices per subject. The dataset was partitioned into 973 subjects for training and 199 subjects for evaluation. The FastMRI challenges offer the generating function for the undersampling mask $\mathcal{M}$. To avoid the inclusion of slices with minimal information, the first five slices of each subject were removed. The TC-DiffRecon network model was built using the PyTorch framework and trained utilizing NVIDIA GeForce RTX 3090 GPUs. Specifically, during the training, the dropout was set at 0.3, the diffusion steps at 4000, the learning rate at 0.0001, and the duration of the training was set at 48h.

\subsection{Comparisons}

We compare the proposed TC-DiffRecon model with the RNN model OUCR\cite{c19guo2021over}, which integrates an over-complete CNN to enhance detail recovery capacity, and with the diffusion model-based DiffuseRecon. Additionally, the reconstruction results were compared with zero-fill reconstruction (ZF) and U-Net \cite{c20ronneberger2015u}. As illustrated in Table \ref{tab1}, our method surpasses the others across all metrics. First, we compared them at 6× and 8× AF, where our model demonstrated superior performance. To ascertain the generalizability of our model, that is, the model's performance under different AF, we input undersampled images under 10× and 4× AF to train the model on 6× and 8× AF images. Compared to models that only target a single AF, there is no degradation in the performance of our model. We also achieve superior results compared to the DiffuseRecon model, which is also general.

\begin{table}
\centering
\caption{Comparison with SoTA method. Where the upper and lower evaluation metrics are PSNR and SSIM respectively. we first compare at the 6× and 8× AF. Then the robustness of the model is verified by applying the {6×, 8×} model on {10×, 4×} downsampled inputs.}
\begin{tabular}{c|c|c|c|c} 
\hline
Methods                          & 6×             & 8×             & 8×$\xrightarrow{}$4×          & 6×$\xrightarrow{}$10×          \\ 
\hline
\multirow{2}{*}{ZF}              & 20.79          & 18.37          & 22.89          & 15.07           \\
                                 & 0.482          & 0.414          & 0.549          & 0.349           \\ 
\hline
\multirow{2}{*}{UNet \cite{c20ronneberger2015u}}            & 26.19          & 25.83          & 27.64          & 21.49           \\
                                 & 0.521          & 0.509          & 0.613          & 0.437           \\ 
\hline
\multirow{2}{*}{OUCR\cite{c16peng2022towards}}            & 27.64          & 26.23          & 27.87          & 24.31           \\
                                 & 0.534          & 0.514            & 0.631          & 0.479           \\ 
\hline
\multirow{2}{*}{DiffuseRecon\cite{c19guo2021over}}    & 27.81          & 26.57          & 27.94           & 25.91           \\
                                 & 0.586          & 0.547          & 0.635          & 0.526           \\ 
\hline
\multicolumn{1}{c|}{TC-DiffRecon} & \textbf{28.74} & \textbf{27.33} & \textbf{29.73} & \textbf{26.45}  \\
\multicolumn{1}{c|}{(Our)}           & \textbf{0.635} & \textbf{0.587} & \textbf{0.739} & \textbf{0.557}  \\
\hline
\end{tabular}
\label{tab1}
\end{table}

\subsection{Ablation studies}
In the following section, we assess the impact of the MF-UNet and texture coordination strategy implemented in the TCKG module on overall model performance through a series of ablation experiments. The results of which are displayed in Table \ref{tab2}. When MF-UNet is removed, the evaluation indexes of the reconstruction results under the four AF show different degrees of degradation. This is due to the fact that the MF-UNet module is able to effectively address the inherent shortcomings of U-Net and avoids excessive smoothing in the generated images. The performance metrics of the reconstruction results are also degraded after removing the texture coordination strategy in the TCKG module. For instance, under the 4× AF, the PSNR reduces from 29.73 to 27.56, and the SSIM decreases from 0.739 to 0.618. These findings suggest that the texture incoherence, inevitably introduced by the data consistency projection, negatively impacts the quality of the reconstruction results. However, our texture coordination strategy effectively mitigates this issue.

\begin{table}
\centering
\caption{Comparison of ablation experiments. The second experiment only removed the texture coordination strategy from the TCKG module.}
\begin{tabular}{c|c|c|c|c} 
\hline
Methods                                  & 4×    & 6×    & 8×    & 10×    \\ 
\hline
\multicolumn{1}{c|}{TC-DiffRecon}      & 26.81      & 26.77      & 26.35      & 25.51       \\
\multicolumn{1}{c|}{(w/o MF-UNet)}        & 0.542      & 0.509      & 0.484      & 0.464       \\ 
\hline
\multicolumn{1}{c|}{TC-DiffRecon}      & 27.56      & 27.14      & 25.87      & 25.27       \\
\multicolumn{1}{c|}{(w/o TCKG)}           & 0.618      & 0.568      & 0.524      & 0.502       \\ 
\hline
\multirow{2}{*}{TC-DiffRecon}         & \textbf{29.73} & \textbf{28.74} & \textbf{27.33} & \textbf{26.45}  \\
                                         & \textbf{0.739} & \textbf{0.635} & \textbf{0.587} & \textbf{0.557}  \\
\hline
\end{tabular}
\label{tab2}
\end{table}

\section{Conclusion}
In this paper, we propose TC-DiffRecon, a novel MRI reconstruction model based on the diffusion model. We replace the denoising model in the diffusion model with MF-UNet, which enhances the realism of the reconstruction results through the dynamic modulation of the two feature vectors. This effectively mitigates the inherent issue of the diffusion model generating overly smooth images. By integrating K-space into the backward diffusion process in a texture-coordinated manner, enhancing our model's robustness against varying AF. Additionally, our approach incorporates the C2F sampling method, accelerating the sampling process while further reducing texture incoherence. Experimental data derived from the FastMRI dataset \cite{c17zbontar2018open} confirm that our model outperforms the SoTA approach, delivering superior results compared to models offering similar robustness to AF. 

\section{Compliance with Ethical Standards}
This research study was conducted retrospectively using human subject data made available in open access by FastMRI dataset\cite{c17zbontar2018open}. Ethical approval was not required as confirmed by the license attached with the open access data.
\vspace{-1.0em}

\section{Acknowledgment}
This work was supported by the Zhejiang Provincial Natural Science Foundation of China (No.LY21F020017), Joint Funds
of the Zhejiang Provincial Natural Science Foundation of China (No.U20A20386), GuangDong Basic and Applied Basic Research Foundation  (No.2022A1515110570), Innovation Teams of Youth Innovation in Science, Technology of High Education Institutions of Shandong Province (No.2021KJ088), Shenzhen Science and Technology Program (No.KCXFZ20201221173008022) and National College Student Innovation and Entrepreneurship Training Program (No. 202310336074).
\vspace{-1.0em}

\bibliographystyle{IEEEbib}

\begin{thebibliography}{10}
\providecommand{\url}[1]{#1}
\csname url@samestyle\endcsname
\providecommand{\newblock}{\relax}
\providecommand{\bibinfo}[2]{#2}
\providecommand{\BIBentrySTDinterwordspacing}{\spaceskip=0pt\relax}
\providecommand{\BIBentryALTinterwordstretchfactor}{4}
\providecommand{\BIBentryALTinterwordspacing}{\spaceskip=\fontdimen2\font plus
\BIBentryALTinterwordstretchfactor\fontdimen3\font minus \fontdimen4\font\relax}
\providecommand{\BIBforeignlanguage}[2]{{%
\expandafter\ifx\csname l@#1\endcsname\relax
\typeout{** WARNING: IEEEtran.bst: No hyphenation pattern has been}%
\typeout{** loaded for the language `#1'. Using the pattern for}%
\typeout{** the default language instead.}%
\else
\language=\csname l@#1\endcsname
\fi
#2}}
\providecommand{\BIBdecl}{\relax}
\BIBdecl

\bibitem{c21madore2002using}
B.~Madore, ``Using unfold to remove artifacts in parallel imaging and in partial-fourier imaging,'' \emph{Magnetic Resonance in Medicine: An Official Journal of the International Society for Magnetic Resonance in Medicine}, vol.~48, no.~3, pp. 493--501, 2002.

\bibitem{c22yu2017deep}
S.~Yu, H.~Dong, G.~Yang, G.~Slabaugh, P.~L. Dragotti, X.~Ye, F.~Liu, S.~Arridge, J.~Keegan, D.~Firmin \emph{et~al.}, ``Deep de-aliasing for fast compressive sensing mri,'' \emph{arXiv preprint arXiv:1705.07137}, 2017.

\bibitem{c4wang2023multimodal}
Y.~Wang, T.~Fu, C.~Wu, J.~Xiao, J.~Fan, H.~Song, P.~Liang, and J.~Yang, ``Multimodal registration of ultrasound and mr images using weighted self-similarity structure vector,'' \emph{Computers in Biology and Medicine}, vol. 155, p. 106661, 2023.

\bibitem{c7markova2022global}
V.~Markova, M.~Ronchetti, W.~Wein, O.~Zettinig, and R.~Prevost, ``Global multi-modal 2d/3d registration via local descriptors learning,'' in \emph{International Conference on Medical Image Computing and Computer-Assisted Intervention}.\hskip 1em plus 0.5em minus 0.4em\relax Springer, 2022, pp. 269--279.

\bibitem{c5he2016deep}
K.~He, X.~Zhang, S.~Ren, and J.~Sun, ``Deep residual learning for image recognition,'' in \emph{Proceedings of the IEEE conference on computer vision and pattern recognition}, 2016, pp. 770--778.

\bibitem{c8montana2022towards}
N.~Monta{\~n}a-Brown, J.~Ramalhinho, B.~Koo, M.~Allam, B.~Davidson, K.~Gurusamy, Y.~Hu, and M.~J. Clarkson, ``Towards multi-modal self-supervised video and ultrasound pose estimation for laparoscopic liver surgery,'' in \emph{International Workshop on Advances in Simplifying Medical Ultrasound}.\hskip 1em plus 0.5em minus 0.4em\relax Springer, 2022, pp. 183--192.

\bibitem{c10ho2020denoising}
J.~Ho, A.~Jain, and P.~Abbeel, ``Denoising diffusion probabilistic models,'' \emph{Advances in neural information processing systems}, vol.~33, pp. 6840--6851, 2020.

\bibitem{c11nichol2021improved}
A.~Q. Nichol and P.~Dhariwal, ``Improved denoising diffusion probabilistic models,'' in \emph{International Conference on Machine Learning}.\hskip 1em plus 0.5em minus 0.4em\relax PMLR, 2021, pp. 8162--8171.

\bibitem{c16peng2022towards}
C.~Peng, P.~Guo, S.~K. Zhou, V.~M. Patel, and R.~Chellappa, ``Towards performant and reliable undersampled mr reconstruction via diffusion model sampling,'' in \emph{International Conference on Medical Image Computing and Computer-Assisted Intervention}.\hskip 1em plus 0.5em minus 0.4em\relax Springer, 2022, pp. 623--633.

\bibitem{c14chung2022score}
H.~Chung and J.~C. Ye, ``Score-based diffusion models for accelerated mri,'' \emph{Medical image analysis}, vol.~80, p. 102479, 2022.

\bibitem{c15gungor2023adaptive}
A.~G{\"u}ng{\"o}r, S.~U. Dar, {\c{S}}.~{\"O}zt{\"u}rk, Y.~Korkmaz, H.~A. Bedel, G.~Elmas, M.~Ozbey, and T.~{\c{C}}ukur, ``Adaptive diffusion priors for accelerated mri reconstruction,'' \emph{Medical Image Analysis}, p. 102872, 2023.

\bibitem{c20ronneberger2015u}
O.~Ronneberger, P.~Fischer, and T.~Brox, ``U-net: Convolutional networks for biomedical image segmentation,'' in \emph{Medical Image Computing and Computer-Assisted Intervention--MICCAI 2015: 18th International Conference, Munich, Germany, October 5-9, 2015, Proceedings, Part III 18}.\hskip 1em plus 0.5em minus 0.4em\relax Springer, 2015, pp. 234--241.

\bibitem{c17zbontar2018open}
J.~Zbontar, F.~Knoll, A.~Sriram, T.~Murrell, Z.~Huang, M.~J. Muckley, A.~Defazio, R.~Stern, P.~Johnson \emph{et~al.}, ``An open dataset and benchmarks for accelerated mri,'' \emph{Fastmri}, vol.~65, 2018.

\bibitem{c18si2023freeu}
C.~Si, Z.~Huang, Y.~Jiang, and Z.~Liu, ``Freeu: Free lunch in diffusion u-net,'' \emph{arXiv preprint arXiv:2309.11497}, 2023.

\bibitem{c19guo2021over}
P.~Guo, J.~M.~J. Valanarasu, P.~Wang, J.~Zhou, S.~Jiang, and V.~M. Patel, ``Over-and-under complete convolutional rnn for mri reconstruction,'' in \emph{Medical Image Computing and Computer Assisted Intervention--MICCAI 2021: 24th International Conference, Strasbourg, France, September 27--October 1, 2021, Proceedings, Part VI 24}.\hskip 1em plus 0.5em minus 0.4em\relax Springer, 2021, pp. 13--23.


\end{thebibliography}


\end{document}